\documentclass[journal=jacsat,manuscript=article]{achemso}
\setkeys{acs}{articletitle=true}

\usepackage[version=3]{mhchem} % Formula subscripts using \ce{}
\usepackage[T1]{fontenc}       % Use modern font encodings
\usepackage{float}
\usepackage{graphicx}
\usepackage{subfig}
\usepackage{epsfig}
\usepackage{epstopdf}
\usepackage{enumerate}

\author{Davinder Singh}

\author{Shubhrangshu Dasgupta}

\email{sdasgupta@iitrpr.ac.in}

\affiliation
{Department of Physics, Indian Institute of Technology Ropar, Rupnagar, Punjab - 140001, India}

\title{Coherence and Its Role in Excitation Energy Transfer in Fenna-Matthews-Olson Complex}

\abbreviations{EET}

\begin{document}

%%%%%%%%%%%%%%%%%%%%%%%%%%%%%%%%%%%%%%%%%%%%%%%%%%%%%%%%%%%%%%%%%%%%%

\begin{tocentry}

{\centering
\includegraphics[width = 1.6 in]{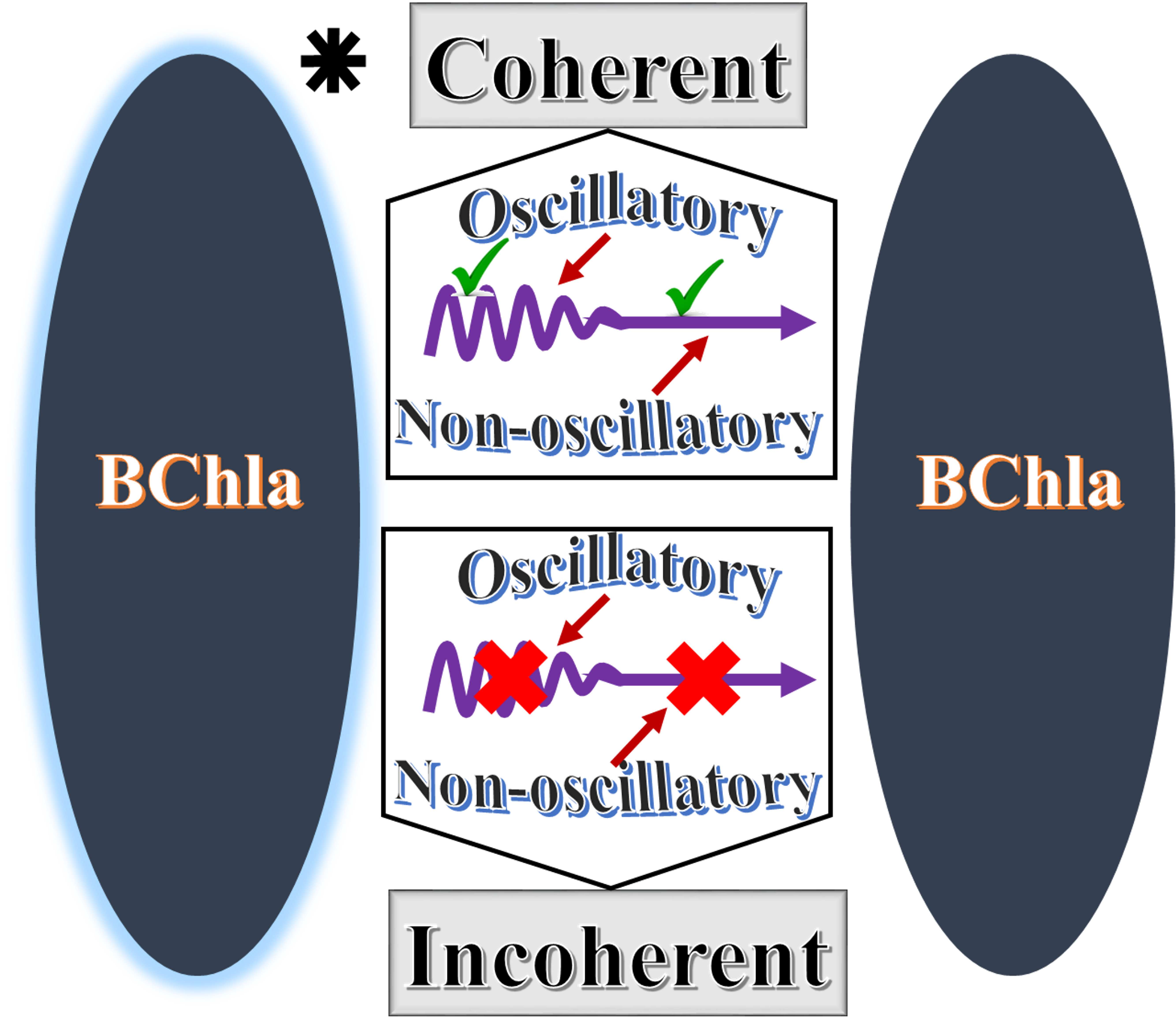}
\par
}
\end{tocentry}

%%%%%%%%%%%%%%%%%%%%%%%%%%%%%%%%%%%%%%%%%%%%%%%%%%%%%%%%%%%%%%%%%%%%%
\begin{abstract}
We show that the coherence between different bacteriochlorophyll-a (BChla) sites in the Fenna-Matthews-Olson complex is an essential ingredient for excitation energy transfer between various sites. The coherence delocalizes the excitation energy, which results in the redistribution of excitation among all the BChla sites in the steady state. We further show that the system remains partially coherent at the steady state. In our numerical simulation of the non-Markovian density matrix equation, we consider both the inhomogeneity of the protein environment and the effect of active vibronic modes.
\end{abstract}

%%%%%%%%%%%%%%%%%%%%%%%%%%%%%%%%%%%%%%%%%%%%%%%%%%%%%%%%%%%%
% \section{Introduction}

Excitation energy transfer (EET) in bacterial photosynthesis is a process in which the excitation moves from one molecule to the other in a very robust fashion on a time scale on the order of a few picoseconds. At the initial stages of photosynthesis,  one of the light-harvesting pigments absorbs solar energy, which is thereafter transferred to the reaction center to store it chemically \cite{Amerongen,Blankenship}. Surprisingly the transfer efficiency of this process is very high, which could be used technologically to meet our future needs of green energy \cite{Meyer_Chemist_2011,Gust_Solar_2009,Maeda_Photocatalytic_2010,
Mallouk_The_2010,Vullev_From_2011,Gust_Mimicking_2010}.

The Fenna-Matthews-Olson (FMO) complex is a pigment-protein complex, which has been extensively studied in last 40 years, in the context of photosynthesis. It transfers the excitation energy from the antenna pigment to the reaction center in green sulfur bacteria through a complex channel of bacteriochlorophyll-a (BChla) molecules \cite{Amerongen,Fenna_Chlorophyll_1975, Sarovar_Quantum_2010}. Such a process had been usually understood in terms of hopping of excitation from one BChla to the other in a unidirectional way. However, in the past decade, experimental observations of EET \cite{Engel_evidence_2007,Panitchayangkoon_long_2010,
panitchayangkoon_direct_2011} using 2D Fourier transform electron spectroscopy \cite{Brixner} of photosynthetic FMO complex has boosted to revisit the idea of quantum effects in photosynthesis. In these experiments, the evidence of oscillations of excitation energy between different BChla sites was interpreted as a reminiscence of quantum coherence. Furthermore, the disappearance of such oscillations was explained in terms of loss of the coherence. It appears that the coherence exists only for the time scale during which the oscillations persist, but such a time scale is a few percent of the total transfer time of excitation energy in FMO complex \cite{Adolphs,chin_noise-assisted_2010}. This leads to an important question: Does the EET remain coherent, even after the oscillation of excitation ceases to exist? We show that the answer is affirmative. Although numerous theoretical approaches have been proposed to understand the role of coherence in EET \cite{Ishizaki,prior_efficient_2010,Pachon,chin_role_2013,
 tiwari_electronic_2013,chenu_enhancement_2013}, it still remains elusive to what extent the coherence affects the EET process\cite{lambert_quantum_2013,scholes_lessons_2011}. We demonstrate the role of coherence in this context through explicit  detailed quantum dynamical simulation. We further show that the strong Coulomb coupling between different BChla sites does not necessarily lead to a large coherence between them. This implies that the coherence plays an {\it independent role} in EET.

\begin{figure}[!h]
\begin{center}
\includegraphics[width = 5 in]{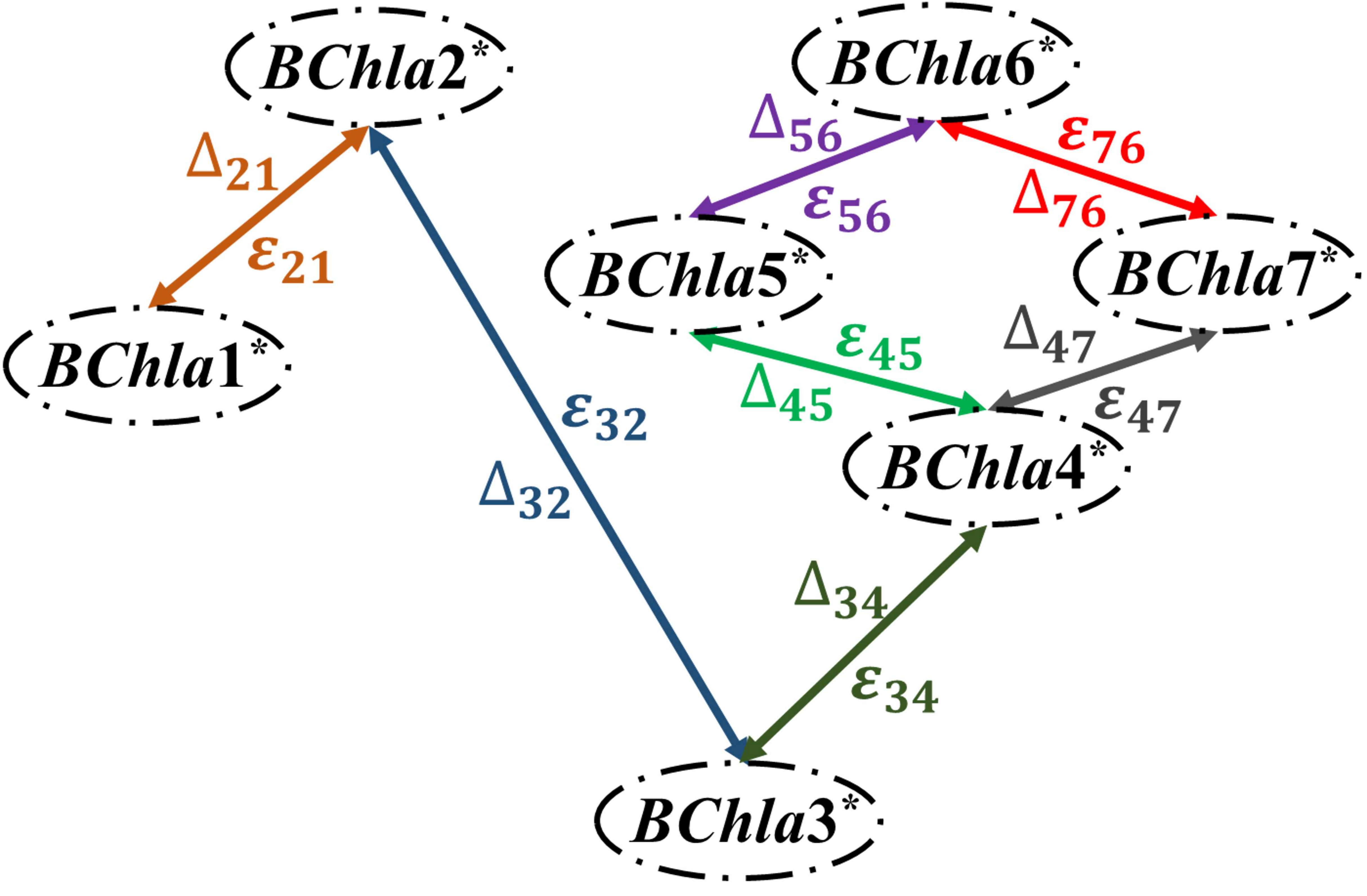}
\caption{(Color online) Schematic illustration of different BChla sites and the dominant channels of EET in a monomer of the FMO complex. The superscript '$\star$' indicates that the molecules are in excited state. Here $\epsilon_{ij}$ represents the energy difference between the transition frequencies of the $i$th and the $j$th BChla site and $\Delta_{ij}$ is the tunnelling frequency between them.}
\label{Levels}
\end{center}
\end{figure}

%%%%%%%%%%%%%%%%%%%%%%%%%%%%%%%%%%%%%%%%%%%%%%%%%%%%%%%%%%%%%%%%
% \section{Model}
FMO complex is a trimer, consisting of three identical monomers \cite{Amerongen,Fenna_Chlorophyll_1975}. Because the Coulomb coupling between the BChla sites of two adjacent monomers is very small, the excitation energy is not shared by these monomers.  Rather the excitation is transferred almost independently through each monomer \cite{Ishizaki,Adolphs}. Each monomer of FMO complex is composed of seven BChla molecules surrounded by the protein molecules \cite{Amerongen,Fenna_Chlorophyll_1975,milder_revisiting_2010}. To analyse the dynamics of EET, each BChla site is modelled as a two-level system. In the subspace of a singly excited BChla molecules, the total Hamiltonian can be written as\cite{Pachon,Gilmore}
\begin{equation}
H = H_{S} + H_{B} + H_{SB}\;,
\label{Hamiltonian}
\end{equation}
where the system Hamiltonian $H_{S}$ is given by
\begin{equation}
H_{S} = \sum_{i,j}\left( \frac{\hbar}{2}\epsilon_{ij}\sigma_{z}^{ij} + \hbar\Delta_{ij}\sigma_{x}^{ij}\right)\;.
\label{System_Hamiltonian}
\end{equation}
Here $\sigma_{z}^{ij}$ and $\sigma_{x}^{ij}$ are the usual Pauli spin matrices relevant to the transitions between the $i$th and the $j$th BChla site (see  Fig. \ref{Levels}), $\epsilon_{ij}$ represents the energy difference between the transition frequencies of these BChla sites, and $\Delta_{ij}$ is the tunnelling frequency between them. Note that this Hamiltonian does not represent a linear Ising chain rather it represents a complex network of different BChla sites as shown in Fig. \ref{Levels}. 

The relevant bath Hamiltonian can be written as
\begin{equation}
H_{B} = \sum\limits_{k_{ij}} \hbar \omega_{k_{ij}}b_{k_{ij}}^\dagger b_{k_{ij}}\;,
\label{Bath_Hamiltonian}
\end{equation}
where, $b_{k_{ij}}$ and $b_{k_{ij}}^\dagger$ are the annihilation and the creation operators, respectively, for the $k$th bath mode, relevant to the transition between the $i$th and $j$th BChla site.
The interaction between the system and the bath can be represented by the standard spin-boson Hamiltonian as
\begin{equation}
H_{SB} = \frac{\hbar}{2}\sum_{i,j} \sum\limits_{k_{ij}} \sigma_{z}^{ij} g_{k_{ij}}(b_{k_{ij}} + b_{k_{ij}}^\dagger) \;,
\label{Interaction_Hamiltonian}
\end{equation}
where $g_{k_{ij}}$ is the electron-phonon coupling constant. 

We employ the density matrix approach to describe the dynamics of the monomer of the FMO complex. 
With the use of the Hamiltonian in Eq. (\ref{Hamiltonian}), we obtain the following master equation, in the Schrodinger picture and in the {\it non-Markovian} limit\cite{Carmichael,Ali}:
\begin{equation}
\begin{array}{lll}
\dot{\rho} &=& -\frac{i}{\hbar}\left[ H_{S},\rho\right]\\
&+&\frac{1}{4}\sum_{i,j}\left\{\left( \sigma_{z}^{ij}\rho\sigma_{z}^{ij} - \rho\sigma_{z}^{ij}\sigma_{z}^{ij}\right)D_{ij}(t)\right. \\
&+&\left( \sigma_{z}^{ij}\rho\sigma_{z}^{ij} - \sigma_{z}^{ij}\sigma_{z}^{ij}\rho\right)D_{ij}^{\ast}(t)\\
&+&\left( \sigma_{z}^{ij}\rho\sigma_{z}^{ij} - \sigma_{z}^{ij}\sigma_{z}^{ij}\rho\right)U_{ij}(t)\\
&+&\left.\left( \sigma_{z}^{ij}\rho\sigma_{z}^{ij} - \rho\sigma_{z}^{ij}\sigma_{z}^{ij}\right)U_{ij}^{\ast}(t) \right\} \;.
\end{array}
\label{non-Markovian}
\end{equation}
The time-dependent coefficients, which contain the information about system-bath correlations, for the transition between the $i$th and the $j$th BChla's are given by
%\small
\begin{eqnarray}
\label{correlation1}D_{ij}(t) &=& \int_{0}^{t}dt^{\prime}\int_{0}^{\infty} d\omega J_{ij}(\omega)\bar{n}(\omega,T)e^{-i\omega(t-t^{\prime})}\;,\\
\label{correlation2}U_{ij}(t) &=& \int_{0}^{t}dt^{\prime}\int_{0}^{\infty} d\omega J_{ij}(\omega)[\bar{n}(\omega,T) + 1]e^{-i\omega(t-t^{\prime})}\;,
\end{eqnarray}
%\normalsize
%\small
%\begin{eqnarray}
%
%
%\end{equation}
%\normalsize
where $\bar{n}(\omega,T)$ is the average number of phonons at an angular frequency $\omega$ in the bath, which is at thermal equilibrium at a temperature $T$ and  $J_{ij}(\omega)$  represents the spectral density function. Several forms, including Ohmic\cite{Pachon} and Lorentzian\cite{Ishizaki}, of this function have been used in previous theoretical studies \cite{adam_on_2013}. Here we choose a much more complex spectral function, as follows, that has been obtained by fitting with the experimental data (obtained using fluorescence line narrowing spectroscopy by Wendling {\it et al.\/}\cite{wendling_electronvibrational_2000})
\begin{equation}
J_{ij}(\omega) = K_{ij}\omega \left( \frac{\omega}{\omega_{c_{ij}}}\right)^{-1/2} e^{-\frac{\omega}{\omega_{c_{ij}}}} + \sum_l K_{l} e^{-\frac{(\omega - \omega_{l})^{2}}{2d^{2}}}\;.
\label{Spectral-Density}
\end{equation}
The above form of spectral density consists of the contributions of vibrational motion, arising from, for example,  the environmental phonons with the Huang-Rhys factor $K_{ij}$ (with $K_{ij} = g_{k_{ij}}^{2}$) and the vibronic modes with the Huang-Rhys factor $K_{l}$. Here $\omega_{c_{ij}}$ is the cutoff frequency and $\omega_l$ represents the frequencies of the active vibronic modes, with dominant Franck-Condon factors. In our analysis, we choose $\omega_{l}$ = 36 $cm^{-1}$, 70 $cm^{-1}$, 173 $cm^{-1}$, 185 $cm^{-1}$, and 195 $cm^{-1}$,  the values of $K_{l}$ equal to 40 times that of the corresponding Franck-Condon factors\cite{wendling_electronvibrational_2000}, and the width of the vibronic band as $d^2 = 18$.

From the structural studies of FMO complex\cite{Amerongen,milder_revisiting_2010,rivera}, it is known that the local protein environment is different for different BChla sites. These differences must be included in the spectral density to suitably model the dynamics of EET. Here this environmental asymmetry is incorporated with the use of different values of $K_{ij}$ and $\omega_{c_{ij}}$ \cite{Davinder_Influence_2016}. Furthermore, spatial arrangement of BChla sites inside a monomer has been explored experimentally\cite{Wen}. It is observed that the BChla 1 and the BChla 6 are close to the baseplate; therefore these two BChlas work as linker pigments to collect excitation from the baseplate. These imply the existence of two pathways of EET,  the first one originating from the BChla 1 and the second one from the BChla 6, as also theoretically suggested\cite{Ishizaki} and as observed in related experiments\cite{Brixner2,cho}. Both the pathways lead the excitation to the  BChla 3 and BChla 4, which are the linker pigments to the reaction center. To simulate the EET dynamics, we use  the values of the site energies and the intersite couplings for the trimeric structure of FMO complex of $Chlorobium$ $tepidum$ as mentioned by Adolphs and Renger \cite{Adolphs}. 

\begin{figure}[!h]
\begin{center}
\subfloat{\label{Ppr2_Bchla1_77_ppltn_chrnc}
\includegraphics[width = 2.0 in]{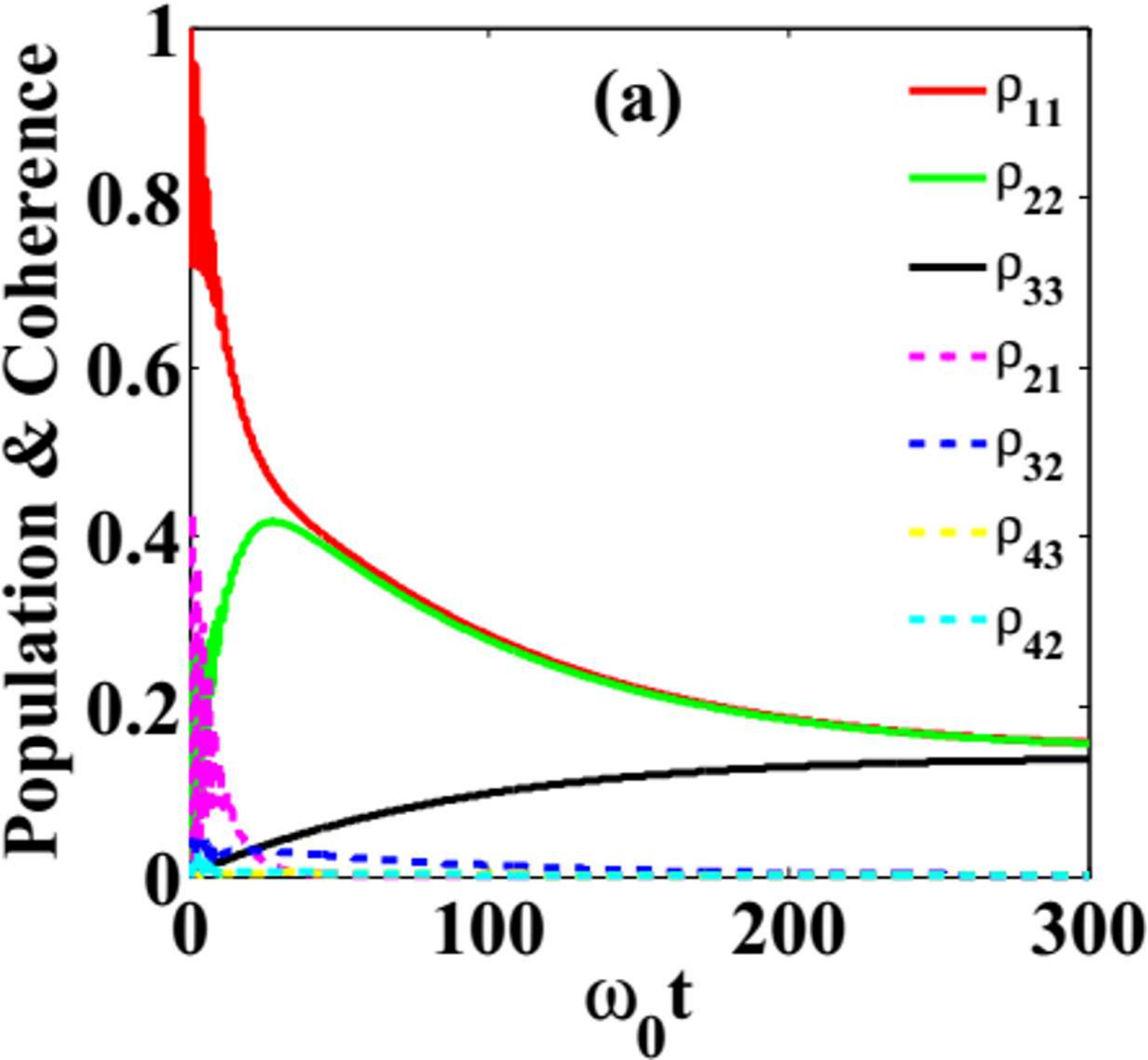}}
\quad
\subfloat{\label{Ppr2_Bchla6_77_ppltn_chrnc}
\includegraphics[width = 1.9 in]{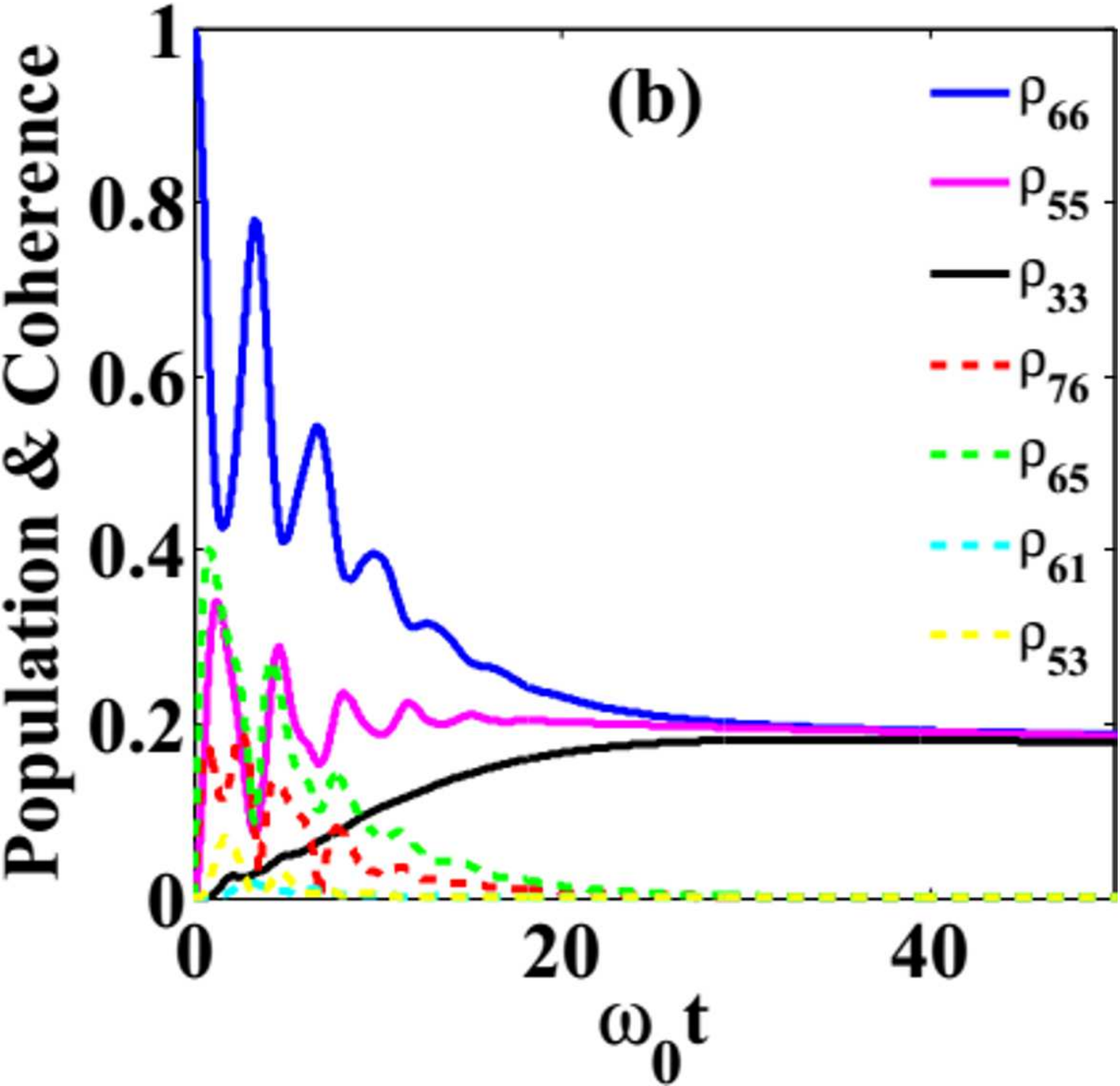}}
\quad
\subfloat{\label{Ppr2_Bchla1_6_77_ppltn_chrnc}
\includegraphics[width = 2.0 in]{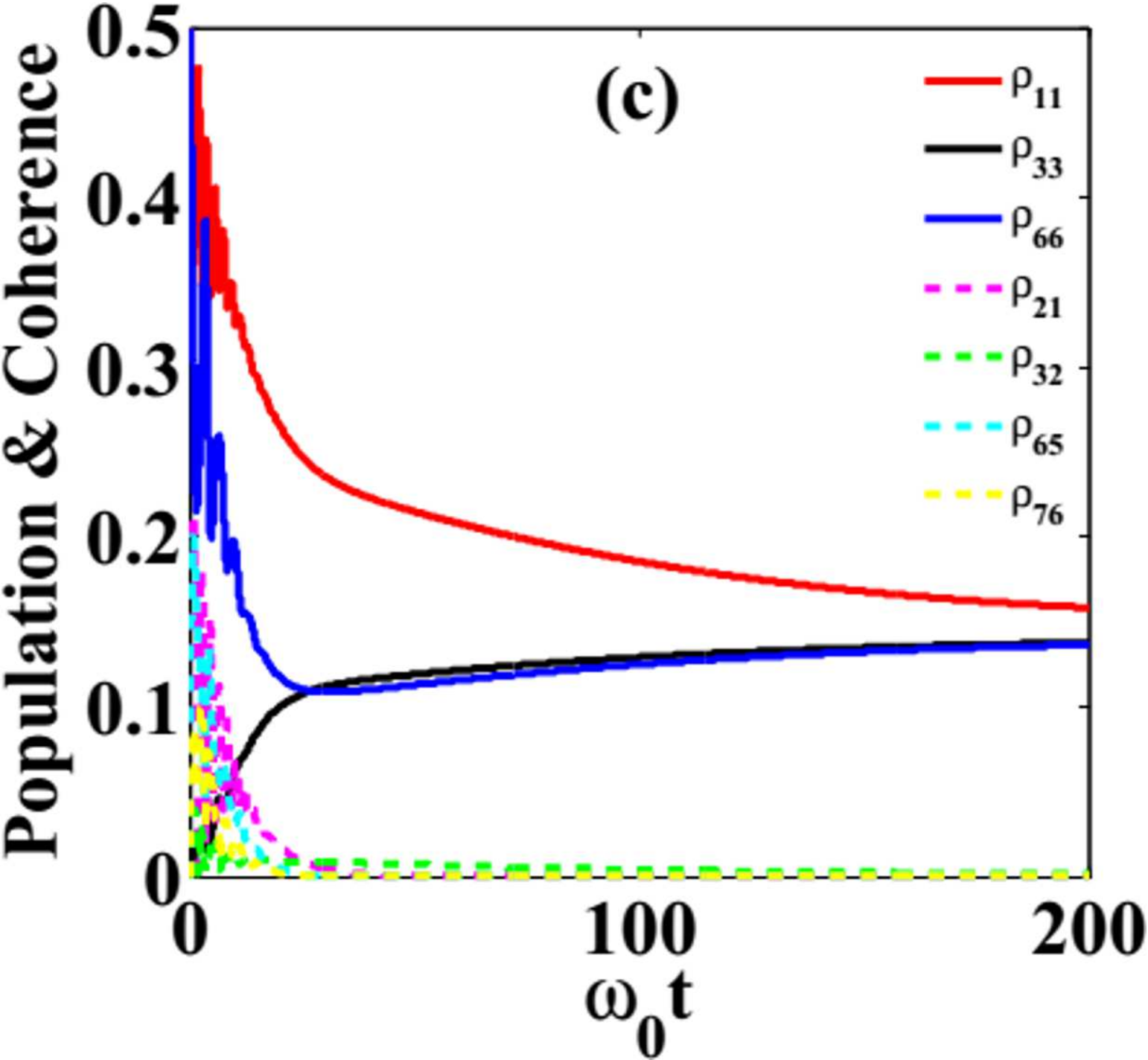}}
\caption{(Color online) Temporal evolution of excitation energy and coherence at cryogenic temperature $T$ = 77 K for the initial excitation at (a) BChla 1, (b)  BChla 6, and (c) both BChla 1 and BChla 6 (equal coherent distribution). We choose $\omega_{0}$ = 100 cm$^{-1}$ for normalisation.}
\label{Ppr2_77_ppltn_chrnc}
\end{center}
\end{figure}

With all of the parameters  as mentioned above, we next numerically solve the non-Markovian Eq. (\ref{non-Markovian}) for the density matrix elements for the two pathways. Note that, the diagonal elements of the density matrix $\rho_{ii}$ represent the 'population', that is, the probability that the $i$th BChla site is excited, while the off-diagonal elements $\rho_{ij}$ ($i \neq j$) describe the coherence between the $i$th and the $j$th BChla sites.  We find that the population in the different BChla sites attain the steady-state values on a time scale on the order of 17 ps (corresponding to $\omega_0t\sim 320$) [see Fig. \ref{Ppr2_Bchla1_77_ppltn_chrnc}], for the initial excitation in BChla 1, while for the initial excitation in the BChla 6, it occurs on a time scale $\sim$2.4 ps (corresponding to $\omega_0 t\sim 45$), as shown in Fig. \ref{Ppr2_Bchla6_77_ppltn_chrnc}. Clearly, the second pathway is much more efficient in transferring population to the BChla 3 and BChla 4. For an equal initial coherent excitation in BChla 1 and BChla 6, the steady state is attained, however,  on a moderate time-scale $\sim$12.7 ps (corresponding to $\omega_0t\sim 240$),  as evident from Fig. \ref{Ppr2_Bchla1_6_77_ppltn_chrnc}.  Note that the excitation oscillates just for $\sim 1$ ps ($\omega_0 t = 18.85$). Such a long time-scale of attaining the steady state, as compared with the time scale of oscillations of excitation, has already been inferred with the use of the modified scaled hierarchical equation of motion approach \cite{zhu_modified_2011}.

\begin{figure}[!h]
\begin{center}
\includegraphics[width = 2.0 in]{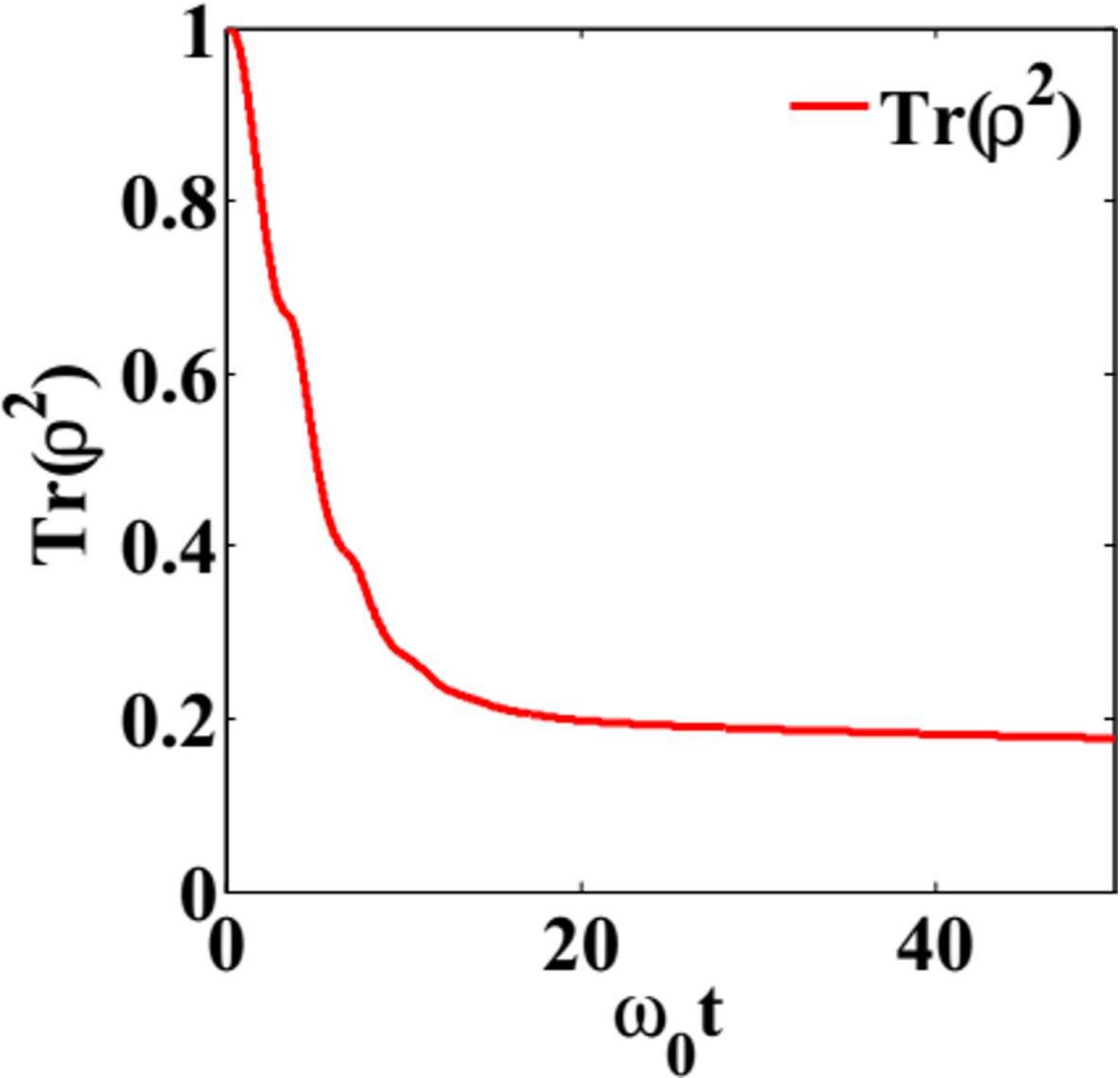}
\caption{(Color online) Temporal evolution of the Tr$(\rho^2)$ at cryogenic temperature $T$ = 77 K for the initial excitation at the BChla 6.}
\label{Trace}
\end{center}
\end{figure}

Usually, in the dynamics of FMO complex, the population oscillations are considered as the signature of coherence. This means that the coherence should be lost as soon as such oscillation does not exist any more. However, our dynamical simulation reveals that the coherence persists for a much longer time scale as compared with the experimentally observed\cite{Engel_evidence_2007,Panitchayangkoon_long_2010,
panitchayangkoon_direct_2011} time scale of oscillations of excitation energy, which is on the order of $\sim 1$ ps (as also evident in Fig. \ref{Ppr2_77_ppltn_chrnc}). 
Hence it is not valid to say that the coherence exists only for the time-scales of population oscillation; rather it also exists for the time during which the population does not oscillate any more. We show in Fig. \ref{Trace} the dynamics of coherence, defined by the Tr$(\rho^2)$, to demonstrate that the coherence indeed persists at the steady state and the system remains partially coherent.  This is different from the Markovian dynamics of a two-level system coupled to a bosonic bath, in which coherence decays much faster than the population attains a steady state value\cite{leggett_dynamics_1987}. Note that for a dimer system, that is composed of only two BChla sites, long time-scales of oscillations of coherence as compared with the time scale of oscillations of excitation are already reported \cite{chin_role_2013}. On the contrary, we observe the existence of both the oscillatory (the `dynamical coherence') as well as the nonoscillatory (the `stationary coherence') time-scales of coherence \cite{mancal_excitation_2013}.

Furthermore, as obvious from the Fig. \ref{Ppr2_77_ppltn_chrnc}, the excitation energy gets distributed among various BChla sites and not only at the target BChla sites, that is, BChla 3 and BChla 4. This is contrary to the other theoretical reports\cite{Ishizaki,zhu_modified_2011}, which show that the excitation is trapped only at the BChla 3 and BChla 4. We emphasize that it is justified in the present model. As the coherence (both the dynamical coherence as well as stationary coherence) delocalizes the excitation over various BChla sites, the excitation energy gets distributed accordingly. The excitation also escapes from the energetic minima. Because of this delocalization, all the BChla sites get an equal amount of excitation energy at the steady state and the trapping of excitation energy at any particular BChla site may not be possible, as obvious from our detailed numerical studies.

\begin{figure}[!h]
\begin{center}

\subfloat{\label{Ppr2_Bchla6_125_ppltn_fully_chrnt}
\includegraphics[width = 2.0 in]{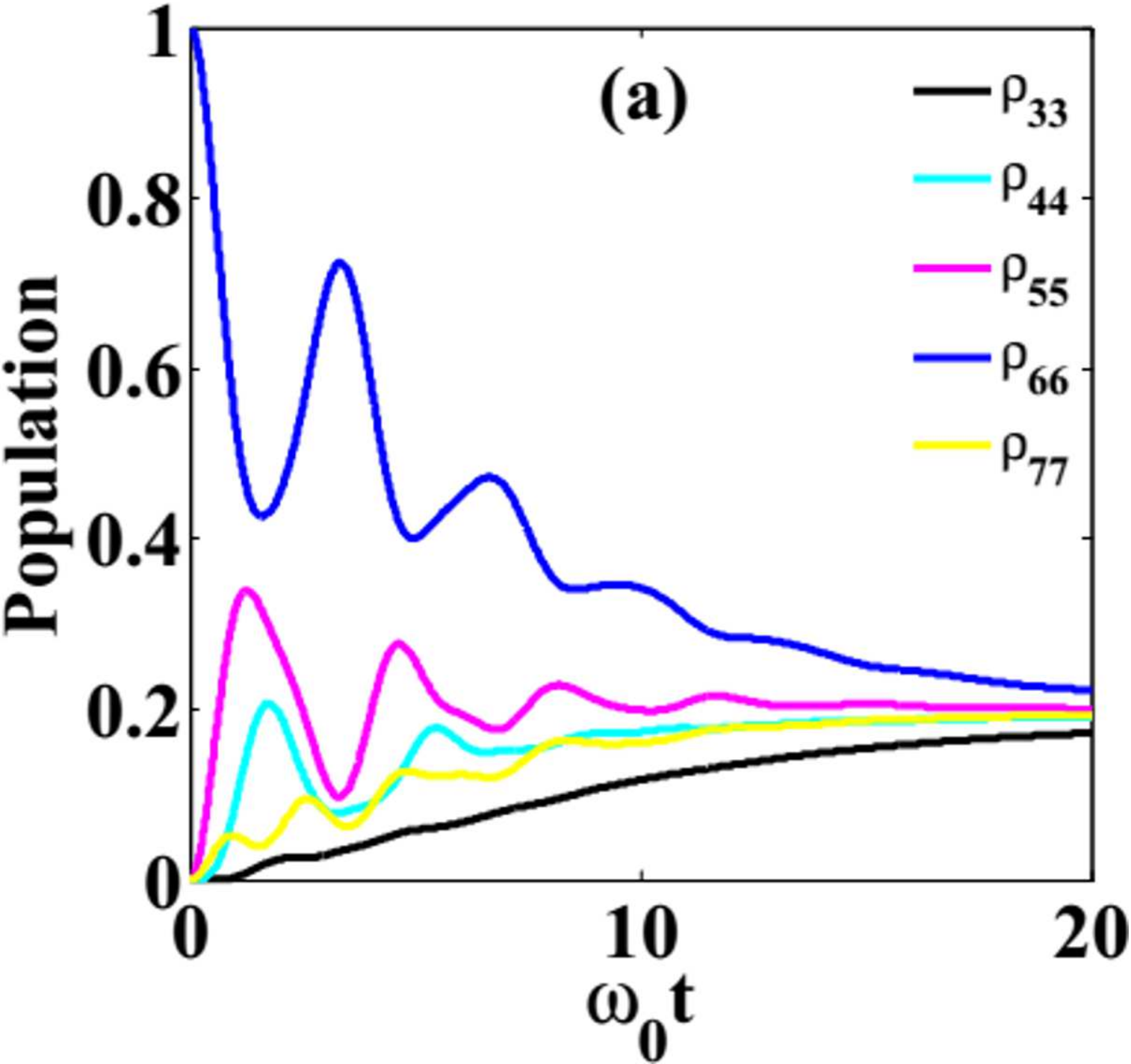}}
\quad
\subfloat{\label{Ppr2_Bchla6_125_chrnc_fully_chrnt}
\includegraphics[width = 2.0 in]{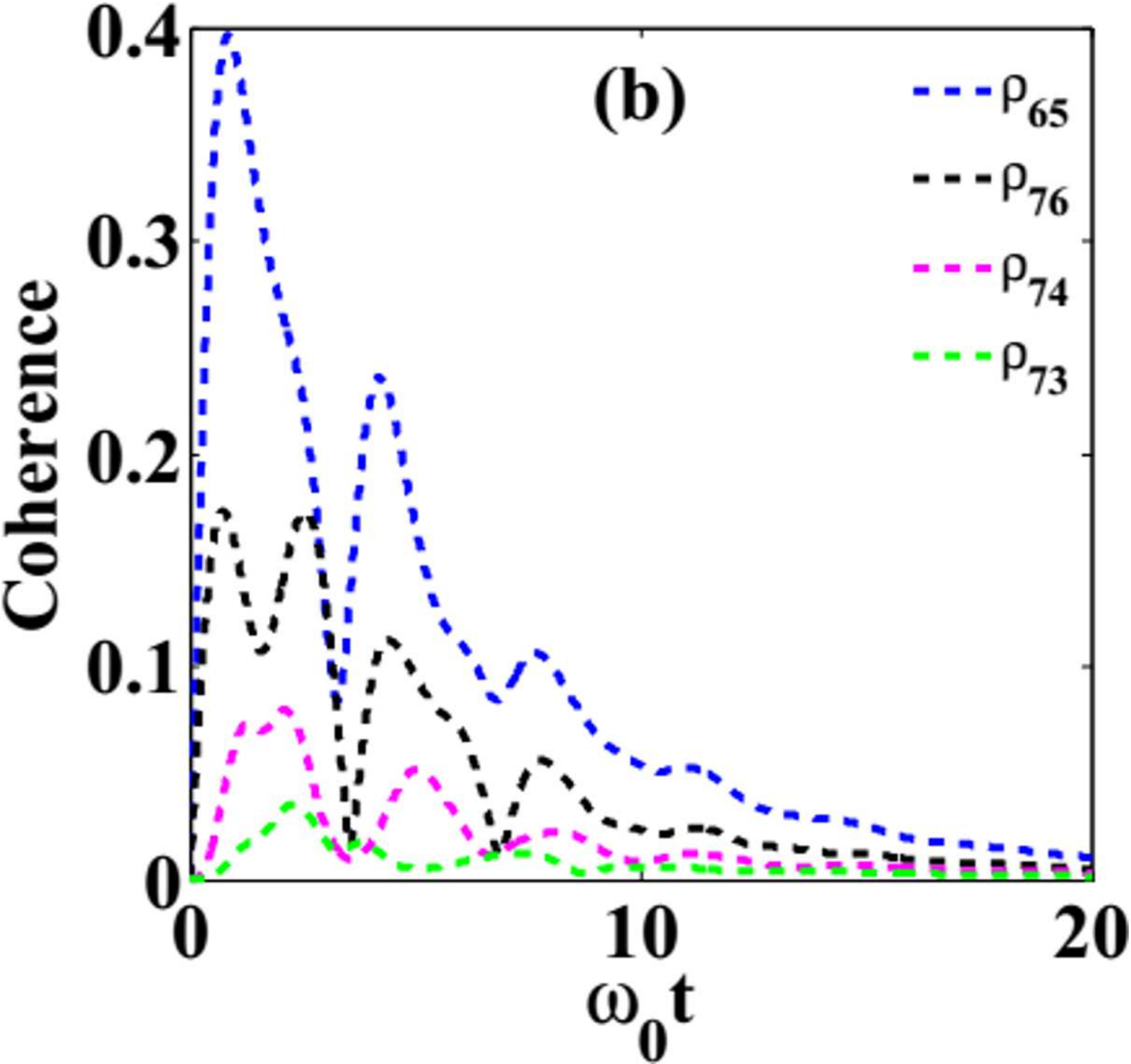}}
\quad
\subfloat{\label{Ppr2_Bchla6_125_chrnc1_fully_chrnt}
\includegraphics[width = 2.0 in]{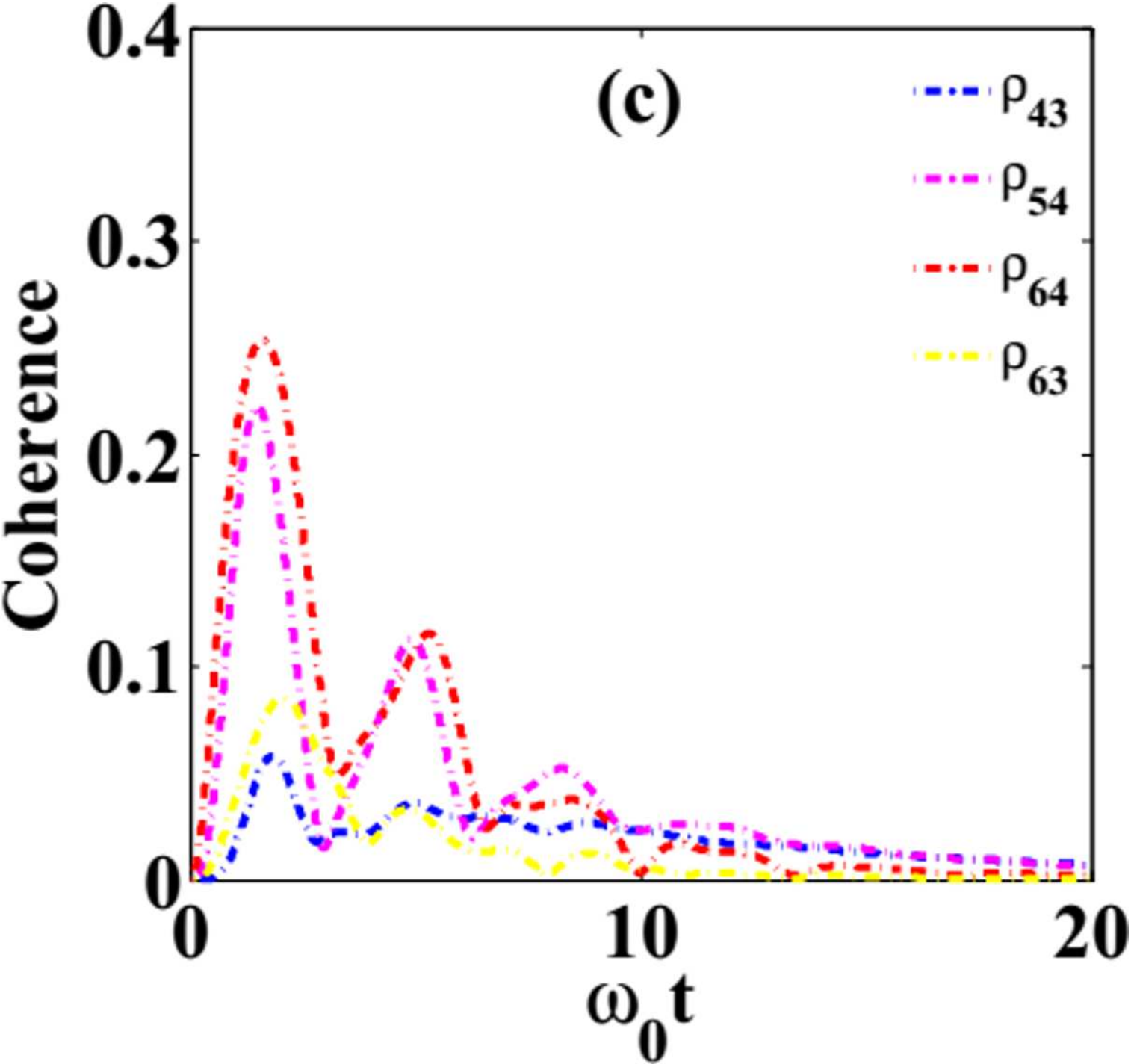}}
\caption{(Color online) Excitation energy transfer dynamics of the second pathway at a temperature $T$ = 125 K. Here (a) represents the population dynamics, while (b) and (c) illustrate the temporal evolution of the coherence between different BChla sites.}
\label{Ppr2_Bchla6_125_fully_chrnt}
\end{center}
\end{figure}

\begin{figure}[!h]
\begin{center}

\subfloat{\label{Ppr2_Bchla6_125_ppltn_par_inchrnt}
\includegraphics[width = 2.0 in]{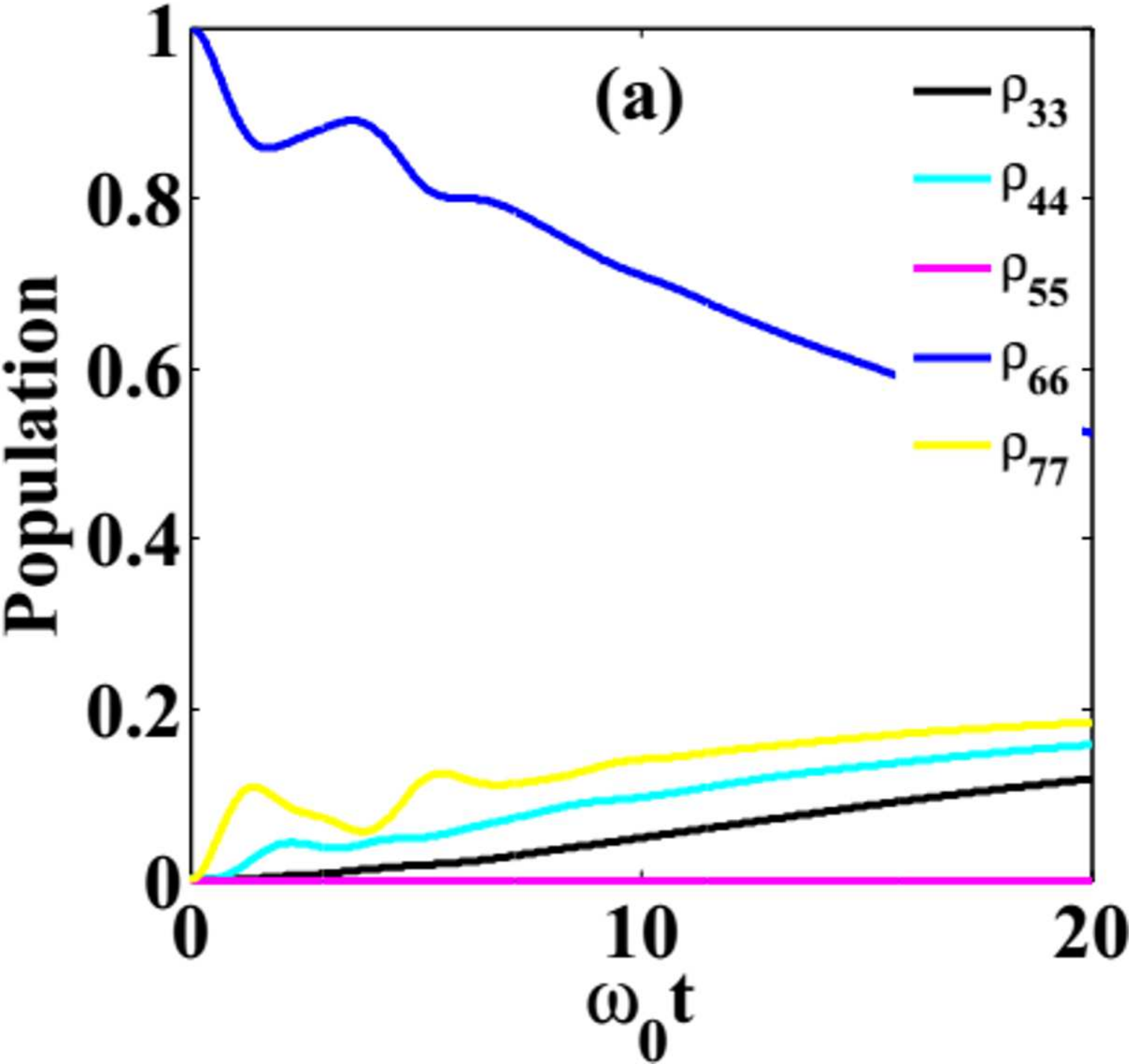}}
\quad
\subfloat{\label{Ppr2_Bchla6_125_chrnc_par_inchrnt}
\includegraphics[width = 2.0 in]{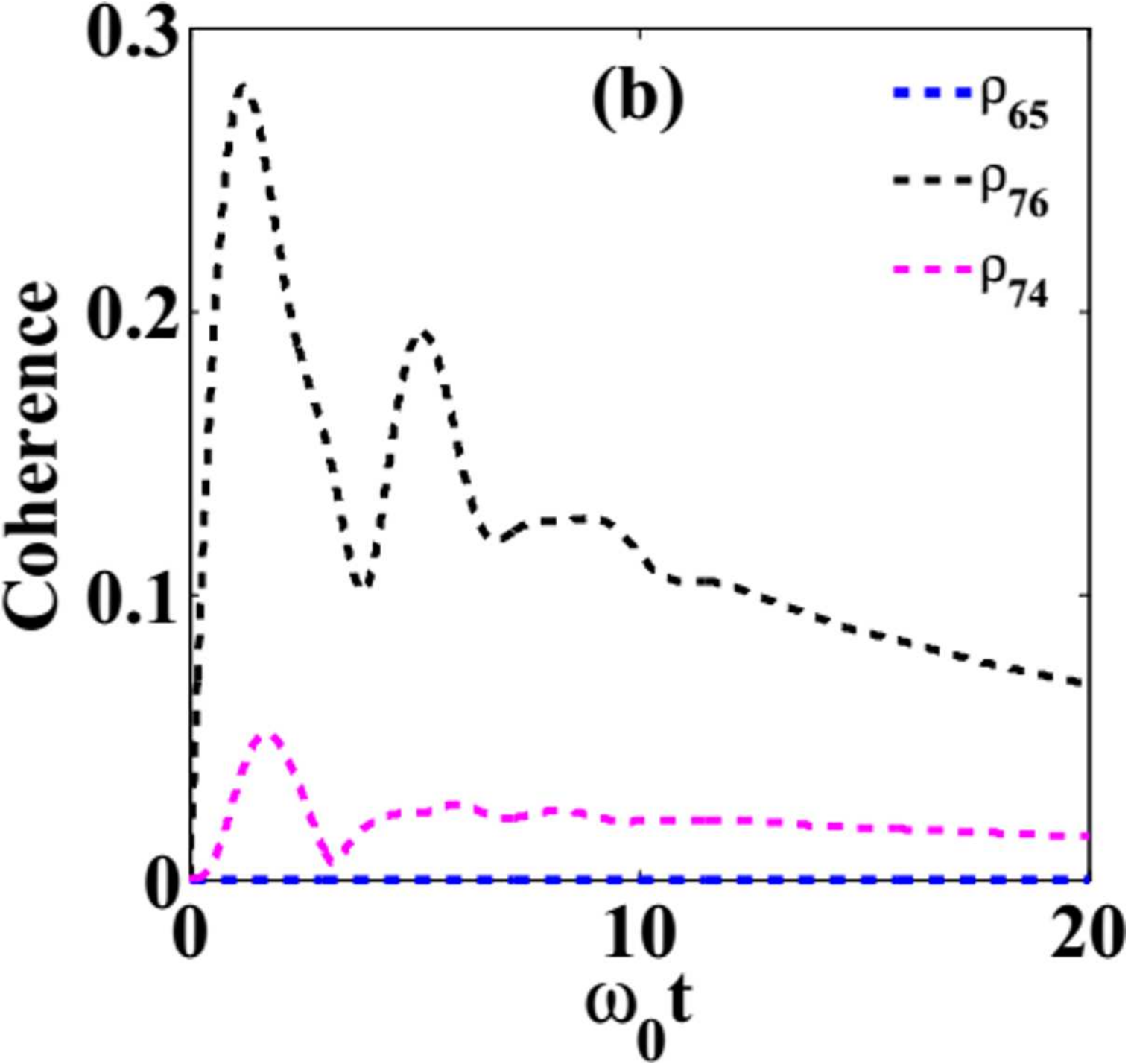}}
\quad
\subfloat{\label{Ppr2_Bchla6_125_chrnc1_par_inchrnt}
\includegraphics[width = 2.0 in]{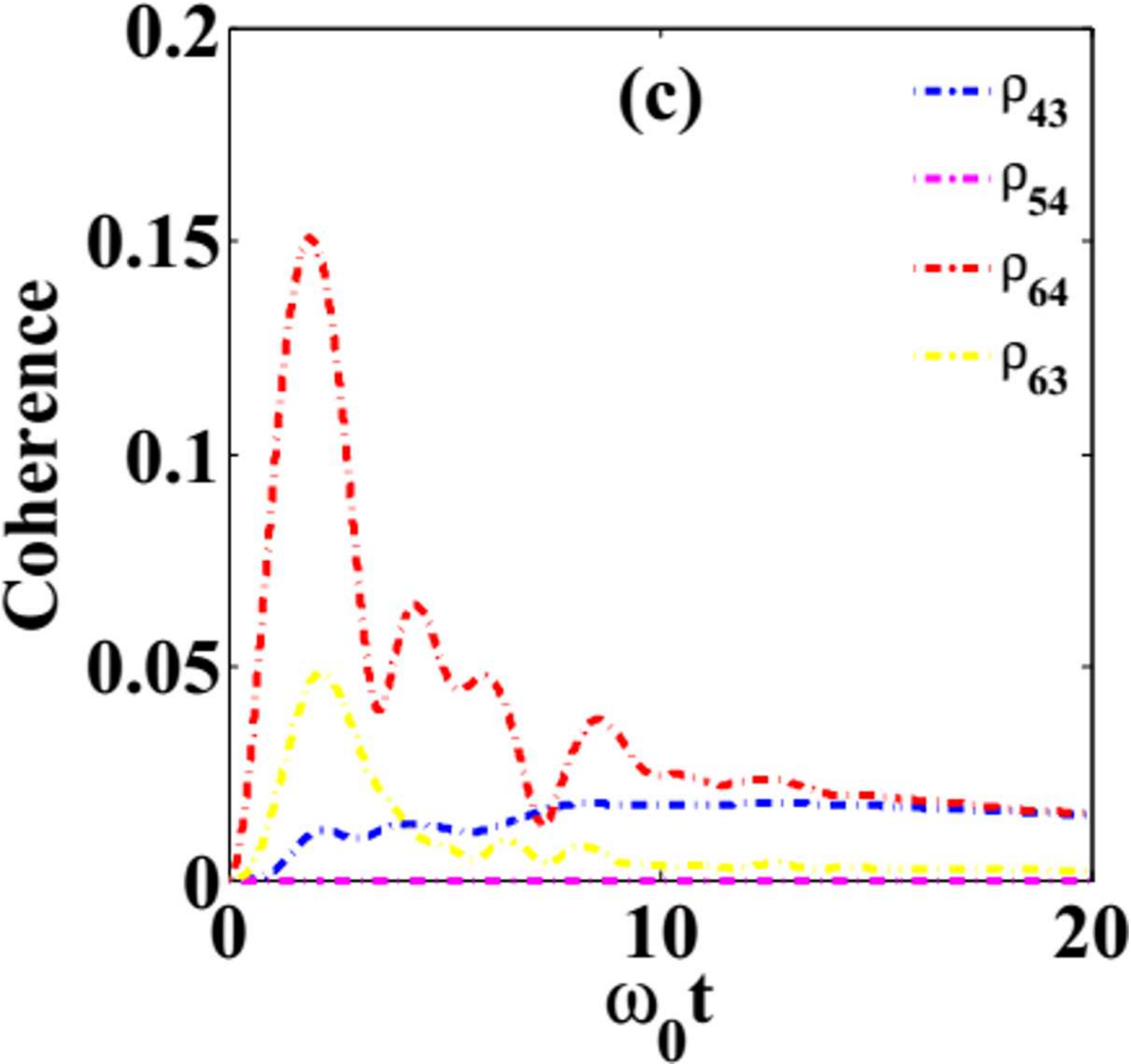}}
\caption{(Color online) Temporal evolution of EET at cryogenic temperature $T$ = 125 K for the second pathway, when it is assumed that BChla 5 is incoherently coupled to the other BChla sites. Here (a) presents the temporal evolution of population of different BChla sites, while (b) and (c) show the dynamics of coherence.}
\label{Ppr2_Bchla6_125_par_inchrnt}
\end{center}
\end{figure}

To further analyse the explicit role of coherence in delocalization of excitation energy, we compare the dynamics of the FMO complex with the nonzero values of all of the coherences and with incoherent coupling to certain BChla sites. In the Fig. \ref{Ppr2_Bchla6_125_fully_chrnt}, all the BChla sites are coherently coupled to each other, while in Fig. \ref{Ppr2_Bchla6_125_par_inchrnt} we assume that BChla 5 is incoherently coupled to the other BChla sites. We model this by choosing the vanishing coherence terms $\rho_{5\alpha}$ ($\alpha \ne 5$) during the entire evolution. In the Fig. \ref{Ppr2_Bchla6_125_ppltn_par_inchrnt}, we  show that there is no transfer of excitation from any BChla site to the BChla 5. This is because, as all the coherence components for the BChla 5 are zero (as illustrated in Fig. \ref{Ppr2_Bchla6_125_chrnc_par_inchrnt} and \ref{Ppr2_Bchla6_125_chrnc1_par_inchrnt}),  the BChla 5 does not get included in delocalization with the other BChla sites. It implies that coherence and hence the delocalization is an essential ingredient for population redistribution among various BChlas. Furthermore, choosing all of the off-diagonal elements of the density matrix to be zero, we find that there is no transfer of excitation to any BChla site, and the entire initial excitation remains with the initial BChla 6 (plots are not shown here). Similar results could be obtained for the initial excitation at the BChla 1.

Note that although the pigment-pigment coupling is an essential ingredient in the EET, the necessity of coherence is not essentially linked to such coupling. In Fig. \ref{Ppr2_Bchla6_125_par_inchrnt}, the dynamics of EET is computed for the nonzero values of pigment-pigment couplings between the BChla 5 and the other BChla sites. We observe that even in the presence of pigment-pigment couplings, the absence of coherence prohibits the EET.  The following cases can be identified in this regard:
\begin{enumerate}[(i)]
\item The pigment-pigment coupling between BChla 6 and BChla 4 is much less than that between BChla 4 and BChla 3, i.e., $\Delta_{64} < \Delta_{43}$, but the coherence between BChla 6 and BChla 4 is often much more than the coherence between BChla 4 and BChla 3, i.e., $\rho_{64}>\rho_{43}$ during the evolution as illustrated in Fig. \ref{Ppr2_Bchla6_125_chrnc1_fully_chrnt}.
\item Both the pigment-pigment coupling as well as the coherence between BChla 6 and BChla 5 are more than those between BChla 7 and BChla 3, i.e., $\Delta_{65} > \Delta_{73}$ as well as $\rho_{65}>\rho_{73}$ as illustrated in Fig. \ref{Ppr2_Bchla6_125_chrnc_fully_chrnt}.
\end{enumerate} 
It essentially implies that the coherence plays an equally important and an independent role as played by the pigment-pigment coupling.

%%%%%%%%%%%%%%%%%%%%%%%%%%%%%%%%%%%%%%%%%%%%%%%%%%%%%%%%%%%%%%%%%%%%%
%\section{Conclusions}

We conclude that in the EET dynamics, the coherence between the different BChla sites is an essential ingredient for the success of EET. The coherence leads to delocalization of the excitation energy across the participating BChla sites and thereby leads to the excitation transfer. The coherence plays an important and an independent role, equally as the pigment-pigment coupling. 

%%%%%%%%%%%%%%%%%%%%%%%%%%%%%%%%%%%%%%%%%%%%%%%%%%%%%%%%%%%%%%%%%%%%%

\begin{acknowledgement}

This work was supported by Department of Science and Technology (DST), Govt. of India, under the grant number SR/S2/LOP-0021/2012.

\end{acknowledgement}

%%%%%%%%%%%%%%%%%%%%%%%%%%%%%%%%%%%%%%%%%%%%%%%%%%%%%%%%%%%%%%%%%%%%%

\bibliography{Paper_2}

\end{document}